\documentclass[12pt]{iopart}
\usepackage{latexsym}
\usepackage{amscd}
\begin{document}

 \def\lambdabar{\protect\@lambdabar}
\def\@lambdabar{%
\relax
\bgroup
\def\@tempa{\hbox{\raise.73\ht0
\hbox to0pt{\kern.25\wd0\vrule width.5\wd0
height.1pt depth.1pt\hss}\box0}}%
\mathchoice{\setbox0\hbox{$\displaystyle\lambda$}\@tempa}%
{\setbox0\hbox{$\textstyle\lambda$}\@tempa}%
{\setbox0\hbox{$\scriptstyle\lambda$}\@tempa}%
{\setbox0\hbox{$\scriptscriptstyle\lambda$}\@tempa}%
\egroup
}

\def\bbox#1{%
\relax\ifmmode
\mathchoice
{{\hbox{\boldmath$\displaystyle#1$}}}%
{{\hbox{\boldmath$\textstyle#1$}}}%
{{\hbox{\boldmath$\scriptstyle#1$}}}%
{{\hbox{\boldmath$\scriptscriptstyle#1$}}}%
\else
\mbox{#1}%
\fi
}
\def\msf{\hbox{{\sf M}}}
\def\msft{\bbox{{\sf M}}}
\def\psf{\hbox{{\sf P}}}
\def\psft{\bbox{{\sf P}}}
\def\Nsf{\hbox{{\sf N}}}
\def\Nsft{\bbox{{\sf N}}}
\def\Tsf{\hbox{{\sf T}}}
\def\Tsft{\bbox{{\sf T}}}
\def\Asft{\bbox{{\sf A}}}
\def\Bsf{\hbox{{\sf B}}}
\def\Bsft{\bbox{{\sf B}}}
\def\Lsf{\hbox{{\sf L}}}
\def\Lsft{\bbox{{\sf L}}}
\def\Ssf{\hbox{{\sf S}}}
\def\Ssft{\bbox{{\sf S}}}
\def\Mtens{\bi{M}}
\def\msfsim{\bbox{{\sf M}}_{\scriptstyle\rm(sym)}}
\newcommand{\mcsim}{ {\sf M}_{ {\scriptstyle \rm {(sym)} } i_1\dots i_n}}
\newcommand{\mcs}{ {\sf M}_{ {\scriptstyle \rm {(sym)} } i_1i_2i_3}}

\newcommand{\beqan}{\begin{eqnarray*}}
\newcommand{\eeqan}{\end{eqnarray*}}
\newcommand{\beqa}{\begin{eqnarray}}
\newcommand{\eeqa}{\end{eqnarray}}

 \newcommand{\suml}{\sum\limits}
\newcommand{\intl}{\int\limits}
\newcommand{\rvec}{\bbox{r}}
\newcommand{\xivec}{\bbox{\xi}}
\newcommand{\Avec}{\bbox{A}}
\newcommand{\Rvec}{\bbox{R}}
\newcommand{\Evec}{\bbox{E}}
\newcommand{\Bvec}{\bbox{B}}
\newcommand{\Svec}{\bbox{S}}
\newcommand{\avec}{\bbox{a}}
\newcommand{\nablav}{\bbox{\nabla}}
\newcommand{\nuvec}{\bbox{\nu}}
\newcommand{\bvec}{\bbox{\beta}}
\newcommand{\vvec}{\bbox{v}}
\newcommand{\jvec}{\bbox{j}}
\newcommand{\nvec}{\bbox{n}}
\newcommand{\pvec}{\bbox{p}}
\newcommand{\mvec}{\bbox{m}}
\newcommand{\evec}{\bbox{e}}
\newcommand{\eps}{\varepsilon}
\newcommand{\la}{\lambda}
\newcommand{\rad}{\mbox{\footnotesize rad}}
\newcommand{\scr}{\scriptstyle}
\newcommand{\latens}{\bbox{\Lambda}}
\newcommand{\pitens}{\bbox{\Pi}}
\newcommand{\cm}{{\cal M}}
\newcommand{\cp}{{\cal P}}
\newcommand{\beq}{\begin{equation}}
\newcommand{\eeq}{\end{equation}}
\newcommand{\ptens}{\bbox{\sf{P}}}
\newcommand{\Ptens}{\bbox{P}}
\newcommand{\Ttens}{\bbox{\sf{T}}}
\newcommand{\Ntens}{\bbox{\sf{N}}}
\newcommand{\Ncal}{\bbox{{\cal N}}}
\newcommand{\Atens}{\bbox{\sf{A}}}
\newcommand{\Btens}{\bbox{\sf{B}}}
\newcommand{\dom}{\mathcal{D}}
\newcommand{\sym}{\scriptstyle \rm{(sym)}}
\newcommand{\Tcal}{\bbox{{\mathcal T}}}
\newcommand{\Nmc}{{\mathcal N}}
\renewcommand{\d}{\partial}
\def\rmi{{\rm i}}
\def\rme{\hbox{\rm e}}
\def\rmd{\hbox{\rm d}}
\newcommand{\ct}{\mbox{\Huge{.}}}
\newcommand{\Laop}{\bbox{\Lambda}}
\newcommand{\Ssfs}{{\scriptstyle \Ssft^{(n)}}}
\newcommand{\Lsfs}{{\scriptstyle \Lsft^{(n)}}}
\title{ A Formula for  Gauge Invariant Reduction of Electromagnetic Multipole Tensors}
\author{C. Vrejoiu}
\address{Facultatea de Fizic\v{a}, Universitatea din Bucure\c{s}ti, 077 125, 
Bucure\c{s}ti-M\v{a}gurele, Rom\^{a}nia,  E-mail :  cvrejoiu@yahoo.com   }
\begin{abstract}
Based on some previous results, one gives a general formula for introducing 
electromagnetic multipole expansions in terms of symmetric and traceless Cartesian 
tensors.  
\end{abstract}

\section{Introduction}
Some advantages of the cartesian forms for multipole moments in the traditional 
formulation of the electromagnetic theory are well known   but the procedure 
of obtaining multipole  tensors corresponding to irreducible  representations of the 
3-dimensional rotation group, especially in the dynamic case, was somehow neglected. 
However, the method done in \cite{ap} for obtaining the symmetric and traceless part of an 
$n$th-rank tensor may be successfully used for this aim.  
     
Let us consider charge $\rho(\rvec,t)$ and current $ \jvec(\rvec,t)$  distributions
having supports included in a finite  domain  ${\cal D}$. Choosing the origin $O$ of
the Cartesian coordinates in $\dom$, the retarded  vector and scalar potentials 
at a point outside ${\cal D}$,  $\rvec=x_i\evec_i$,  are given by the the multipolar 
expansions

\beqa\label{dezv1}
&~&\Avec(\rvec,t)=\frac{\mu_0}{4\pi}\nablav\times\suml_{n\geq 1}\frac{(-1)^{n-1}}
{n!}\nablav^{n-1}\ct\left[\frac{1}{r}\msft^{(n)}(t-r/c)\right]\nonumber\\
&~&+\frac{\mu_0}{4\pi}\suml_{n\geq 1}\frac{(-1)^{n-1}}
{n!}\nablav^{n-1}\ct\left[\frac{1}{r}\dot{\psft}^{(n)}(t-r/c)\right]\nonumber\\
&~&\Phi(\rvec,t)=\frac{1}{4\pi\eps_0}\suml_{\geq 0}\frac{(-1)^n}{n!}\nablav^n\ct
\left[\frac{1}{r}\psft^{(n)}(t-r/c)\right]
\eeqa
where, denoting  by ${\Ttens}^{(n)}$ a $n$th-order tensor, ${\Atens}^{(n)}\ct{\Btens}^{(m)}$
 is a $|n-m|$th-order tensor with the components:
\beqan
\fl&~& ({\Atens}^{(n)}\ct{\Btens}^{(m)})_{i_1 \cdots i_{|n-m|}}=
\left\{\begin{array}{ll}
A_{i_1 \cdots i_{n-m}j_1 \cdots j_m}B_{j_1 \cdots j_m} & \;\rm{,}\; n>m\\
A_{j_1 \cdots j_n}B_{j_1 \cdots j_n} & \;\rm{,}\; n=m\\
A_{j_1 \cdots j_n}B_{j_1 \cdots j_n i_1 \cdots i_{m-n}} & \;\rm{,}\; n<m
\end{array} \right.
\eeqan
The electric and magnetic moments are defined by
\beqa\label{defpm}
\fl&~&\psft^{(n)}(t)=\intl_{\dom}\xivec^n\rho(\xivec,t)\rmd^3\xi:\;
\psf_{i_1\dots i_n}=\intl_{\dom}\xi_{i_1}\dots\xi_{i_n}\rho(\xivec,t)\rmd^3\xi\nonumber\\
\fl&~&\msft^{(n)}(t)=\frac{n}{n+1}\intl_{\dom}\xivec^n\times\jvec(\xivec,t)\rmd^3\xi:\;\;
\msf_{i_1\dots i_n}=\frac{n}{n+1}\intl_{\dom}\xi_{i_1}\dots \xi_{i_{n-1}}(\xivec\times\jvec)_{i_n}
\rmd^3\xi
\eeqa
As one sees, the $\psft^{(n)}$ electric tensor is fully symmetric; the magnetic tensor 
$\msft^{(n)}$ is symmetric in  the first $n-1$ indices and
\beq\label{ctm}
\msf_{i_1\dots i_{k-1}\,j\,i_{k+1}\dots i_{n-1}\,j}=0,\; k=1\dots i_{n-1}.
\eeq 
 It is shown in \cite{vr}  that we can introduce  such transformations of the multipole 
 tensors
\beq\label{red}
\psft^{(n)}\longrightarrow \widetilde{\psft}^{(n)},\;\;
\msft^{(n)}\longrightarrow \widetilde{\msft}^{(n)},
\eeq
where $\widetilde{\psft}^{(n)},\;\widetilde{\msft}^{(n)}$ are fully symmetric and traceless tensors, 
so that,  if $\widetilde{\Avec}$ and $\widetilde{\Phi}$ are obtained from equations 
\eref{dezv1} by the substitutions \eref{red}, then
$$\widetilde{\Avec}(\rvec,t)=\Avec(\rvec,t)+\nablav\Psi(\rvec,t),\;
\widetilde{\Phi}(\rvec,t)=\Phi(\rvec,t)-\frac{\d}{\d t}\Psi(\rvec,t),\;\;
\Box\Psi=0$$
the correspondence $(\Avec,\,\Phi)\longrightarrow (\widetilde{\Avec},\,\widetilde{\Phi})$ 
being a gauge transformation.
\par The present paper is an attempt to systematize some results of the author and of the 
 co-workers in this field resuming them in a compact formula.   

\section{A Gauge Invariant Procedure of Reducing Multipole Tensors}
 We present bellow the results from \cite{vr} by a way initiated in \cite{DV}.
\footnote{In \cite{DV} the demonstrations are given using the charge and current densities 
expansions}
In  the transformations \eref{red}  the operations of obtaining the symmetric 
and traceless part of some tensors are implied. Generally, we will have to obtain the symmetric parts  
of   $n$th-rank tensors of magnetic type, i.e. being symmetric in the first $n-1$ indices 
and verifying the property \eref{ctm}. Let $\Lsft^{(n)}$ such a tensor. Then, the symmetric 
part of this tensor is given by
\beq\label{simL}
\fl \Lsf_{\sym i_1\dots i_n}=\frac{1}{n}\big[\Lsf_{i_1\dots i_n}+
\Lsf_{i_ni_2\dots i_1}+\dots +\Lsf_{i_1\dots i_n\,i_{n-1}}\big]
=\Lsf_{i_1\dots i_n}-\frac{1}{n}\suml^{n-1}_{\la=1}\eps_{i_{\la}i_nq}\Ncal^{(\la)}
_{i_1\dots i_{n-1}\,q}\big[\Lsft^{(n)}\big]
\eeq
where $\Nmc^{\dots (\la)}_{\dots}$ is the component with the index $i_{\la}$ suppressed.
\par The operator $\Ncal$ introduced in equation \eref{simL} defines a correspondence 
 between $\Lsft^{(n)}$ and an $n-1$th-rank tensor: 
\beq\label{opN}
\fl\Lsft^{(n)}\longrightarrow \Ncal\big[\Lsft^{(n)}\big]:\;
 \left[\Ncal\big[\Lsft^{(n)}\big]\right]_{i_1\dots i_{n-1}}\equiv 
 \Nmc_{i_1\dots i_{n-1}}\big[\Lsft^{(n)}\big]=
 \eps_{i_{n-1}ps}\Lsf_{i_1\dots i_{n-2}ps}
 \eeq 
which is  fully symmetric in the first $n-2$ indices and the contractions of the last 
index with the precedent indices give null results. So, the tensor $\Ncal[\Lsft^{(n)}]$ 
is  of the type $\msft^{(n-1)}$. 
Particularly,
\beqa\label{ms1}
\fl&~&\Nmc_{i_1\dots i_{n-1}}\big[\msft^{(n)}\big]=\frac{n}{n+1}
\intl_{\dom}\xi_{i_1}\dots\xi_{i_{n-2}}\big[\xivec\times
(\xivec\times\jvec)\big]_{i_n}\rmd^3\xi,\nonumber\\
\fl&~&\left[\Ncal^{\,2}\big[\msft^{(n)}\big]\right]_{i_1\dots i_{n-2}}\equiv
\Nmc^{\,2}_{i_1\dots i_{n-2}}\big[\msft^{(n)}\big]=-\frac{n}{n+1}
\intl_{\dom}\xi^2\xi_{i_1}\dots\xi_{n-3}(\xivec\times\jvec)_{i_{n-2}}\rmd^3\xi
\eeqa
and it is easy to see that
\beqa\label{ms2}
\fl&~&\Ncal^{\,2k}\big[\msft^{(n)}\big]=\frac{(-1)^nn}{n+1}\intl_{\dom}\xi^{2k}
\xivec^{n-2k}\times\jvec\rmd^3\xi,\nonumber\\
\fl&~&\Ncal^{\,2k+1}\big[\msft^{(n)}\big]=\frac{(-1)^kn}{n+1}\intl_{\dom}
\xi^{2k}\xivec^{n-2k-1}\times(\xivec\times\jvec)\rmd^3\xi,\;k=0,1,2\dots
\eeqa
Let a fully symmetric tensor $\Ssft^{(n)}$ and the {\it detracer} operator $\Tcal$  
introduced in \cite{ap}. This operator acts on a totally symnmetric tensor $\Ssft^{(n)}$ 
so that $\Tcal[\Ssft^{(n)}]$ is a fully symmetric and traceless tensor of rank $n$. 
The {\it detracer theorem} of Applequist \cite{ap} states that
\beq\label{detrth}
\fl\left[\Tcal\big[\Ssft\big]\right]_{i_1\dots i_n}=
\suml^{[n/2]}_{m=0}(-1)^m(2n-1-2m)!!\suml_{D(i)}\delta_{i_1i_2}\dots
\delta_{i_{2m-1}i_{2m}}\Ssf^{(n:m)}_{i_{2m+1}\dots i_n}
\eeq
where $[n/2]$ denotes the integer part of $[n/2]$,  the sum over $D(i)$ is the sum 
over all permutations of the indices $i_1\dots i_n$ which give distinct terms and 
$\Ssf^{(n:m)}_{i_{2m+1}\dots i_n}$ denotes the components of the $(n-2m)$th-order tensor 
obtained from $\Ssf^{(n)}$ by contracting $m$ pairs of symbols $i$. 
\par It is usefull to introduce here another operator $\Laop$ by the relation
\beq\label{defla}
 \left[\Tcal\big[\Ssft^{(n)}\big]\right]_{i_1\dots i_n}=
\Ssf_ {i_1\dots i_n}-\suml_{D(i)}\delta_{i_1i_2}\left[\Laop\big[\Ssft^{(n)}\big]\right]
_{i_3\dots i_n}
\eeq
where $\left[\Laop\big[\Ssft^{(n)}\big]\right]
_{i_3\dots i_n}$ define a fully symmetric tensor of rank $n-2$. From this definition 
together with the theorem \eref{detrth}, we obtain
\beq\label{lamatr}
\fl\Lambda_{i_3\dots i_n}\big[\Ssft^{(n)}\big]=\suml^{[n/2]}_{m=1}
\frac{(-1)^{m-1}(2n-1-2m)!!}{(2n-1!!m}\suml_{D(i)}\delta_{i_3i_4}\dots \delta_{i_{2m-1}
i_2m}\Ssf^{(n:m)}_{i_{2m+1}\dots i_n}.
\eeq
In the following, for simplifying the notation, any argument of the operator 
$\Laop$ is considered as a symmetrized tensor i.e. $\Laop[\Tsf^{(n)}]=\Laop[\Tsf^{(n)}
_{\scr sym}]$ for any tensor $\Tsf^{(n)}$. Same observation applies to the operator 
$\Tcal$: $\Tcal[\Tsf^{(n)}]=\Tcal[\Tsf^{(n)}_{\scr sym}]$.
\par The following four transformation properties of the multipole tensors and potentials   
may be used for establishing the results from \cite{vr}.
\par 1- Let the transformation of the $n$th-order magnetic tensor 
\footnote{We suppose that the tensors $\Lsft$  and $\Ssft$, used for 
obtaining general transformation relations, are function of $t-r/c$.} :
\beq\label{trLM}
\fl \msft^{(n)}\rightarrow\msft^{(n)}_{(L)}:\;
\msf_{(L)i_1 \dots i_n}=\msf_{i_1 \dots i_n}-\frac{1}{n}\suml^{n-1}_{\la=1}
\eps_{i_{\la}i_nq}\Nmc^{(\la)}_{i_1 \dots i_{n-1}\,q}\big[\Lsf^{(n)}].
\eeq  
Let us substitute in the expansion of the potential $\Avec$ the tensor $\msft^{(n)}$ by 
$\msft^{(n)}_{(L)}$ obtaining 
\beqa\label{trA2}
\fl&~& \Avec[{\scriptstyle \msft^{(n)}\rightarrow\msft^{(n)}_{(L)}}]=\Avec
-\frac{\mu_0}{4\pi}\frac{(-1)^{n-1}}{n!n}\,\evec_i\,\d_j\d_{i_1}\dots\d_{i_{n-1}}
\left[\frac{1}{r}\suml^{n-1}_{\la=1}\eps_{ijk}\eps_{i_{\la}kq}
\Nmc^{(\la)}_{i_1\dots i_{n-1}q}\big[\Lsft^{(n)}\big]\right]\nonumber\\
\fl&~&=\Avec
+\frac{\mu_0}{4\pi}\frac{(-1)^{n-1}}{n!n}\,\evec_i\,
\d_j\d_{i_1}\dots\d_{i_{n-1}}
\frac{1}{r}\big[\big(\delta_{i\,i_1}\Nmc_{i_2\dots i_{n-1}\,j}+\dots\delta_{i\,i_{n-1}}
\Nmc_{i_1\dots i_{n-2}\,j}\big)\nonumber\\
\fl&~&- \big(\delta_{j\,i_1}\Nmc_{i_2\dots i_{n-1}\,i}+\dots\delta_{j\,i_{n-1}}
\Nmc_{i_1\dots i_{n-2}\,i}\big)\big]\big[\Lsft^{(n)}\big]\nonumber\\
\fl&~&= \Avec+\frac{\mu_0}{4\pi}\frac{(-1)^{n-1}(n-1)}{n!n}\,\evec_i\,\d_{i_1}\dots \d_{i_{n-2}}
\left[\d_j\d_i\left(\frac{1}{r}\Nmc_{i_1\dots i_{n-2}j}\right)
-\d_j\d_j\left(\frac{1}{r}\Nmc_{i_1\dots i_{n-2}i}\right)\right]\nonumber\\
\fl&~&=\Avec+\nablav\Psi(\rvec,t)
+\frac{\mu_0}{4\pi}\frac{(-1)^{n}(n-1)}{n!c^2n}\nablav^{n-2}\ct \left[\frac{1}{r}
\ddot{\Ncal}\big[\Lsft^{(n)}\big]\right].
\eeqa   
Here the relation $\eps_{ijk}\eps_{i_{\la}kq}=\delta_{iq}\delta_{ji_{\la}}-
   \delta_{ii_{\la}}\delta_{jq}$ is considered and, also, that for  $r\,\ne\, 0$
  $$ (\Delta-\frac{1}{c^2}\frac{\d^2}{\d t^2})\frac{f(t-r/c)}{r}=0. $$
  The function $\Psi$ is given by
  $$\Psi(\rvec,t)=\frac{\mu_0}{4\pi}\frac{(-1)^{n-1}(n-1)}{n!n}\nablav^{n-1}\ct\left[\frac{1}{r}
  \Ncal^{(n-1)}(t-r/c)\right],\;\; \Box\Psi(\rvec,t)=0.$$
 Let the transformation
 \beq\label{1}
 \fl
 \psft^{(n-1)}\rightarrow\psft^{'(n-1)}=\psft^{(n-1)}+a_1(n)\dot{\Ncal}
 \big[\Lsf^{(n)}\big],\;\;a_1(n)=-\frac{n-1}{c^2n^2}.
 \eeq
 Introducing the transformed potentials produced by the substitution 
 $ \psft^{(n-1)}\rightarrow\psft^{'(n-1)}$, we obtain 
\beqan
\fl&~& \Avec[{\scriptstyle \msft^{(n)}\rightarrow\msft^{(n)}_{(L)},\;
\psft^{(n-1)}\rightarrow\psft^{'(n-1)}}]=\Avec+\nablav\Psi,\\
\fl&~&\Phi[{\scriptstyle \psft^{(n-1)}\rightarrow\psft^{\,'(n-1)}}]=\Phi
-\frac{\d \Psi}{\d t}.
\eeqan
So, the transformation \eref{trLM} produces changes in the potentials which, up to 
a gauge transformation, are compensated by the transformation \eref{1}.  	
\par 2-Let the transformation of the $n$th-order electric tensor:
\beq\label{trLP}
 \psft^{(n)}\rightarrow\psft^{(n)}_{(L)}:\;
\psf_{(L)i_1 \dots i_n}=\psf_{i_1 \dots i_n}-\frac{1}{n}\suml^{n-1}_{\la=1}
\eps_{i_{\la}i_nq}\Nmc^{(\la)}_{i_1 \dots i_{n-1}\,q}\big[\Lsf^{(n)}].
\eeq   
We obtain
\beqan
\fl&~& \Avec[{\scriptstyle \psft^{(n)}\rightarrow\psft^{(n)}_{(L)}}]=\Avec
+\frac{\mu_0}{4\pi}\frac{(-1)^{n-1}(n-1)}{n!\,n}\nablav\times\left\{\nablav^{n-2}\ct
\left[\frac{1}{r}\dot{\Ncal}\big[\Lsft^{(n)}\big]\right]\right\},\\
\fl&~&\Phi[{\scriptstyle \psft^{(n)}\rightarrow\psft^{(n)}_{(L)}}]=\Phi.
\eeqan
The change of the vector potential  is compensated by the transformation
\beq\label{2}\fl 
\msft^{(n-1)}\longrightarrow \msft^{(n-1)}+a_2(n)\dot{\Ncal}\big[\Lsft^{(n)}\big],\;\;
a_2(n)=\frac{n-1}{n^2}.
\eeq
\par 3-Let the transformation of the magnetic vector of rank $n$:
\beq\label{trsM}
\msft^{(n)}\longrightarrow \msft^{(n)}_{\scriptstyle (S)}:\;\;
\msft_{{\scriptstyle (S)}i_1\dots i_n}=\msft_{i_1\dots i_n}-\suml_{D(i)}\delta_{i_1i_2}
\Lambda_{i_3\dots i_n}\big[\Ssft^{(n)}\big]
\eeq
where $\Ssft^{(n)}$ is a fully symmetric tensor.
\par The change in the vectorial potential produced by this transformation is
\beqan
\fl&~&\Avec[{\scriptstyle \msft^{(n)}\rightarrow\msft^{(n)}_{(S)}}]=
\Avec-\frac{\mu_0}{4\pi}\frac{(-1)^{n-1}(n-2)(n-1)}{2\,n!}\nablav\times \left[
\nablav^{n-3}\ct\bbox{\Lambda}\big[\Ssft^{(n)}\big]\right]
\eeqan
This alteration of the vectorial potential is eliminated by the transformation
\beq\label{3}\fl 
\msft^{(n-2)}\longrightarrow\msft^{(n-2)}+b(n)\ddot{\bbox{\Lambda}}\big[\Ssft^{(n)}\big],
\;\;\;b(n)=\frac{n-2}{2c^2n}.
\eeq 
\par 4-The transformation
\beq\label{trsP}
\psft^{(n)}\longrightarrow \psft_{\scriptstyle (S)}:\;\;
\psft_{{\scriptstyle (S)}i_1\dots i_n}=\psft_{i_1\dots i_n}-\suml_{D(i)}\delta_{i_1i_2}
\Lambda_{i_3\dots i_n}\big[\Ssft^{(n)}\big]
\eeq
produces the following changes of the potentials:
\beqan
\fl&~&\Avec[{\scriptstyle \psft^{(n)}\rightarrow \psft^{(n)}_{(S)}}]=
\Avec-\frac{\mu_0}{4\pi}\frac{(-1)^{n-1}}{n!}\evec_i\d_{i_1}\dots\d_{i_{n-1}}
\left[\frac{1}{r}\suml_{D(i)}\delta_{i_1i_2}\dot{\Lambda}_{i_3\dots i_n}
\big[\Ssft^{(n)}\big]\right]\\
\fl&~&=\Avec+\nablav \Psi'-\frac{\mu_0}{4\pi}\frac{(-1)^{n-1}(n-2)(n-1)}{2n!c^2}
\nablav^{n-3}\ct\left[\frac{1}{r}\tdot{\bbox{\Lambda}}\big[ \Ssft^{(n)}\big]
\right]
\eeqan	
where 
$$\Psi'=
-\frac{\mu_0}{4\pi}\frac{(-1)^{n-1}(n-1)}{n!}\nablav\left[\nablav^{n-2}\ct
\left(\frac{1}{r}\dot{\bbox{\Lambda}}\big[\Ssft^{(n)}\big]\right)\right]
$$
and
\beqan
\fl&~&\Phi\big[{\scriptstyle \psft^{(n)}\rightarrow\psft^{(n)}_{(S)}}\big]=
\Phi+\frac{\mu_0}{4\pi}\frac{(-1)^{n-1}n(n-1)}{2n!}\nablav^{n-2}\ct\left[\frac{1}{r}
\ddot{\bbox{\Lambda}}\big[\Ssft^{(n)}\big]\right].
\eeqan
Let the transformation 
\beq\label{4}\fl 
\psft^{(n-2)}\longrightarrow \psft"^{(n-2)}=\psft^{(n-2)}+b(n)\ddot{\bbox{\Lambda}}
\big[\Ssft\big]
\eeq
with $b(n)$ given by equation \eref{3}. The effect of the transformation \eref{4} 
on the potential $ \Avec$ is the compensation of the extra-gauge term . So
$$\Avec\big[{\scriptstyle \psft^{(n)}\rightarrow\psft^{(n)}_{(S)},
\psft^{(n-2)}\rightarrow\psft^{"(n-2)}}\big]=\Avec+\nablav\Psi'$$
but it is easy to see that the modification of the scalar potential $\Phi$ produced 
by the transformation \eref{4} together with the modification due to the transformation 
\eref{trsP} give
$$\Phi\big[{\scriptstyle \psft^{(n)}\rightarrow\psft^{(n)}[\Ssft^{(n)}],
\psft^{(n-2)}\rightarrow\psft^{"(n-2)}}\big]=\Phi-\frac{\d\Psi'}{\d t}$$
the total effect of the transformations \eref{trsP} and \eref{4} being a gauge transformation 
of the potentials.  
\section{General Formula}
Let the gauge invariant process of reducing the multipole tensor begins for the electric tensors 
from the order $n=\eps$ and for the magnetic ones from $n=\mu$. Generally, as seen, for example, 
from the calculation of the total power radiated by a confined system of charges and currents,  
 $\eps > \mu$. Moreover, in this case, for a consistent consideration of the 
multipolar expansion of the radiated power, it suffices to get $\eps=\mu+1$, \cite{DV}. In the following we consider the cases 
$\eps >\mu$.   
The following formulae are results of the rules represented by equations 
\eref{1}, \eref{2}, \eref{3} and \eref{4}.

\beqa\label{ForP}
\fl\widetilde{\psft}^{(n)}={\mathcal P}^{(n)}&+&{\mathcal T}\left\{
\suml^{[(\eps-n)/2]}_{k=1}A^{(n)}_k\frac{\rmd^{2k}}{\rmd t^{2k}}\Lambda^k\big[
\psft^{(n+2k)}\big]\right.\nonumber\\
\fl&~&+\left.\suml^{[(\mu-n-1)/2]}_{k=0}\frac{\rmd^{2k+1}}{\rmd t^{2k+1}}\suml^k_{l=0}
B^{(n)}_{kl}\latens^l\Ncal^{2k-2l+1}\big[\msft^{n+1+2k}\big]\right\},
\eeqa
\beqa\label{ForM}
\fl&~&\widetilde{\msft}^{(n)}={\mathcal M}^{(n)}+{\mathcal T}\left\{
\suml^{[(\mu-n)/2]}_{k=1}\frac{\rmd^{2k}}{\rmd t^{2k}}\suml^{k}_{l=0}
C^{(n)}_{kl}\latens^l\Ncal^{2k-2l}\big[\msft^{(n+2k)}\big]\right\}
\eeqa
where
\beqa\label{AB}
\fl&~& A^{(n)}_k=\prod^{k}_{l=1}b(n+2l),\nonumber\\
\fl&~&B^{(n)}_{kl}=\prod^l_{q=1}b(n+2q)\prod^{k-l}_{h=0}a_1(n+1+2k-2h)\prod^{k-l-1}
_{s=0}a_2(n+2k-2s)
\eeqa
and
\beqa\label{C}
\fl&~&C^{(n)}_{kl}=\prod^l_{q=1}b(n+2q)\prod^{k-l-1}_{h=0}a_1(n+2k-2h)\prod^{k-l-1}_{s=0}
a_2(n-1+2k-2s).
\eeqa
By ${\mathcal P}^{(n)}$ and ${\mathcal M}^{(n)}$ we understand the "static" expressions of 
the reduced multipole tensors:
\beqan
\fl&~& {\mathcal P}^{(n)}(t)={\mathcal T}[\psft^{(n)}]=\frac{(-1)^n}{(2n-1)!!}\intl_{\mathcal D}
\rho(\rvec,t)r^{2n+1}\nablav^n\frac{1}{r}\rmd^3x,\\
\fl&~&{\mathcal M}^{(n)}(t)={\mathcal T}[\msft^{(n)}]=\frac{(-1)^n}{(n+1)(2n-1)!!}\suml^n_{\la=1}
\intl_{\mathcal D}r^{2n+1}\big[\jvec(\rvec,t)\times\nablav\big]_{i_{\la}}\d^{(\la)}
_{i_1\dots i_n}\frac{1}{r}\rmd^3x.
\eeqan

In these formulae one considers 
$$\prod^L_{k=l} F_k=1 \;\mbox{if}\;L\,<\,l.$$
For justifying the formulae \eref{ForP}, \eref{ForM}, we may consider separately the reduction of the 
tensors $\psft^{(n)}$ beginning from $n=\eps$. Using equation \eref{4}, we obtain
\beq\label{ec1}
\fl\psft^{(\eps-2p)}\rightarrow\widetilde{\psft}^{'(\eps-2p)}={\mathcal P}^{(\eps-2p)}+{\mathcal T}
\left[\suml^p_{k=1}A^{(\eps-2p)}_k \frac{\rmd^{2k}}{\rmd t^{2k}}\latens^k \big[
\psft^{(\eps-2p+2k)}\big]\right],\;\;p=0,1,2,\dots.
\eeq 
Repeating the operations but beginning from $n=\eps-1$, we obtain the full set of 
reduced electric tensors for $\mu=0$.
 The process of reduction is a little more complicated for the magnetic tensors because 
 of the operation of symmetrization. Beginning with $n=\mu$ and applying equations 
 \eref{1}, \eref{2}, we obtain the following results for the symmetrized tensors:
 
 \beq\label{ec2}
 \fl  \msft^{(\mu-2p)}\rightarrow \msft^{(\mu-2p)}_{\scr sym}
 +\left\{\suml^{p}_{k=1}C^{(\mu-2p)}_{k0}\frac{\rmd^{2k}}{\rmd t^{2k}}\Ncal^{2k}\big[
 \msft^{(\mu-2p+2k)}\big]\right\}_{\scr sym},
 \eeq
 \beq\label{ec3}
 \fl\widetilde{\psft}^{'(\mu-2p-1)}\rightarrow \widetilde{\psft}^{'(\mu-2p-1)}+\left\{
 \suml^p_{k=0}B^{(\mu-2p-1)}_{k0}\frac{\rmd^{2k+1}}{\rmd t^{2k+1}}\Ncal^{2k+1}
 \big[\msft^{(\mu-2p+2k)}\big]\right\}_{\scr sym}.
 \eeq
 By applying equations \eref{3} and \eref{4} to these results, we obtain the equations 
 \eref{ForP} and \eref{ForM}. 
 \section{Conclusions}
 Equations  \eref{ForP} and \eref{ForM} show that it is possible  to give compact formulae 
 for the electromagnetic multipolar expansions using the general tensorial formalism 
 and all is reduced to simple algebraic calculations which may be performed also by  
 automatic numerical or symbolic computation.

\vspace{0.5cm}

\par 

{\bf References}
\begin{thebibliography}{10}
\bibitem{ap}
J. Applequist, {\it J. Phys. A: Math. Gen.}, {\bf 22} (1989) 4303-4330

\bibitem{vr}
C. Vrejoiu, {\it J. Phys. A: Math. Gen.}, {\bf 35} (2002) 9911-22
\bibitem{DV}
I. Dumitriu, C. Vrejoiu, {\it Preprint arXiv:physics/0503164 v1}

\end{thebibliography}
\end{document}